# Dispersive wave control enabled by silicon metamaterial waveguides

T. T. D. Dinh[1,*], X. Le Roux[1], M. Montesinos-Ballester[1], C. Lafforgue[1], J. Zhang[1], D. González-Andrade[1], D. Melati[1], D. Bouville[1], D. Benedikovic[2], P. Cheben[3,4], E. Cassan[1], D. Marris-Morini[1], L. Vivien[1], C. Alonso-Ramos[1,*]

[1]Centre de Nanosciences et de Nanotechnologies, CNRS, Université Paris-Saclay, Palaiseau 91120, France
[2]Dept. Multimedia and Information-Communication Technologies, University of Žilina, 01026 Žilina, Slovakia
[3]National Research Council Canada, 1200 Montreal Road, Bldg. M50, Ottawa, Ontario K1A 0R6, Canada
[4]Center for Research in Photonics, University of Ottawa, Ottawa, Ontario K1N 6N5, Canada

*Corresponding author: thi-thuy-duong.dinh@c2n.upsaclay.fr , carlos.ramos@c2n.upsaclay.fr

**The ability to exploit the on-chip nonlinear generation of new frequencies has opened the door to a plethora of applications in fundamental and applied physics. Excitation of dispersive waves is a particularly interesting process that allows efficient nonlinear wavelength conversion by phase-matching of a soliton and waves in the normal dispersion regime. However, controlling the wavelength of dispersive waves in integrated waveguides remains an open challenge, hampering versatile shaping of the nonlinear spectral response. Here, we show that metamaterial silicon waveguides release new degrees of freedom to select the wavelength of dispersive waves. Based on this concept, we experimentally demonstrate excitation of two dispersive waves near 1.55 µm and 7.5 µm allowing ultra-wideband supercontinuum generation, covering most of the silicon transparency window. These results stand as an important milestone for versatile nonlinear frequency generation in silicon chips, with a great potential for applications in sensing, metrology and communications.**

Nonlinear optics is an excellent candidate for the generation of new frequency components and the spectral broadening of light, with applications in sensing, metrology and communications [1]. Recent demonstrations show the great potential of nanophotonic waveguides for on-chip nonlinear optics, as they provide tight mode confinement, with modal areas near 1 µm², that result in a substantial reduction of the power required to trigger nonlinear effects. Excitation of dispersive waves is a particularly appealing process allowing high-efficiency nonlinear wavelength conversion, exploiting phase-matching and frequency overlap of solitons and wavelengths in the normal dispersion regime [1]. Integrated waveguides generally have non negligible high-order dispersion coefficients ($\beta_{k\leq 4} \neq 0$), resulting in the possible generation of two dispersive waves, one on each side of the soliton spectrum. Controlling the wavelength of the dispersive waves could enable ad-hoc efficient wavelength conversion, with great potential for quantum [2] and spectroscopy [3] applications.

The position of the dispersive waves also determines the maximum achievable bandwidth of supercontinuum generation (SCG). Most on-chip SCG demonstrations utilized a similar approach to maximize the bandwidth, based on designing the waveguide cross section to excite dispersive waves with wavelengths located far away from the optical pump wavelength [4–19]. This strategy allowed remarkable SCG demonstrations, including among others, 1 octave SCG in the near-IR with indium gallium phosphide [5], 2.3 octaves SCG in the visible to mid-IR range with silicon nitride [10], 1.94 octaves SCG in the mid-IR range with silicon [14], and 2.1 octaves SCG in the mid-IR range with germanium [17]. Still, proposed strip and rib waveguides yield limited flexibility in the control of the position of dispersive waves, as phase-matching at the two sides of the soliton is strongly affected by variations in the waveguide dimensions, precluding independent selection of the position of the two dispersive waves. Phase matching between different modes in periodically structured waveguides has been proposed to control the position of dispersive waves [12, 20]. However, the efficiency of this approach may be limited by the comparatively weak spatial overlap between different waveguide modes.

In this work, we propose a novel approach to control the excitation of dispersive waves in integrated waveguides, based on silicon metamaterial waveguides. We utilize the additional geometrical degrees of freedom in the metamaterial waveguide to control the position of the two dispersive waves, independently. The metamaterial waveguide geometry, presented in Fig. 1, is composed by a central strip and a lateral metamaterial-grating cladding. The metamaterial cladding, with a period (Λ) shorter than half of the pump wavelength, allows removal of the silica under-cladding, obviating the silica absorption for wavelengths above 4 µm, while providing mechanical stability and effective-low index cladding to confine the optical mode [21–23]. We exploit the membrane configuration and the unique degrees of freedom released by the metamaterial engineering to tune the position of the dispersive waves. To illustrate the potential of this approach, we experimentally demonstrate supercontinuum generation expanding 2.35 octaves between two dispersive waves near 1.55 µm and 7.5 µm wavelength, which covers most of the silicon transparency window.

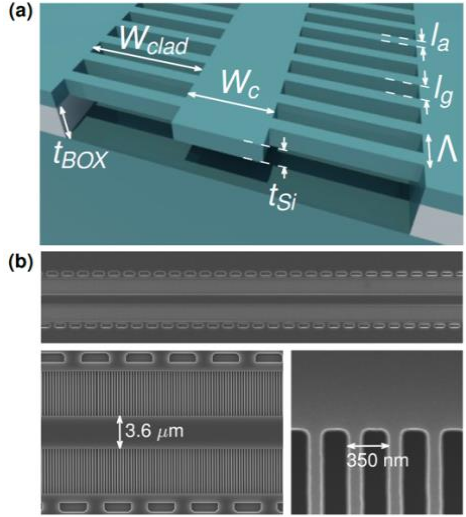

Fig. 1. (a) 3D Schematic view of the dispersion-engineered silicon metamaterial waveguide. (b) Scanning electron microscope image of the fabricated waveguide.

Since their first demonstration in silicon photonics [24, 25], subwavelength-grating metamaterials [26, 27] have been used as a powerful tool for overcoming performance limitations of conventional silicon photonic devices. Previous theoretical studies suggest that subwavelength grating metamaterial structuration has a great potential for dispersion engineering of silicon waveguides [28, 29]. Yet, these works focused on the control of the group velocity dispersion to maximize the bandwidth of the anomalous dispersion but did not analyze dispersive wave excitation. In addition, they considered silicon-on-insulator (SOI) waveguides, and thereby were limited to wavelengths below 4 µm by the absorption of the silica cladding [13, 14].

The position of the dispersive waves can be predicted by solving the phase matching condition [1]:

$$\beta_{int} - \frac{\gamma P}{2} = 0 \qquad (1)$$

$P$ is the peak power of the soliton. The nonlinear coefficient $\gamma$ is determined by $\gamma = (\omega_o n_2(\omega_o))/(c A_{eff}(\omega_o))$, with $\omega_o$ the pulse center frequency, $c$ the light constant, $n_2$ the nonlinear coefficient, and $A_{eff}(\omega_o)$ the effective modal area at $\omega_o$. The integrated dispersion $\beta_{int}$ is defined as

$$\beta_{int} = \beta(\omega_{DW}) - \beta(\omega_S) - \frac{\omega_{DW} - \omega_S}{v_{g,s}}, \qquad (2)$$

being $\omega_{DW}$ the frequency of the dispersive wave, $\omega_S$ the frequencies of most energetic soliton arising from the initial pulse, $v_{g,s}$ the soliton group velocity and $\beta(\omega)$ the propagation constant.

To illustrate the flexibility in controlling the dispersive wave position enabled by our approach, in Fig. 2, we compare the group velocity dispersion (D) and the integrated dispersion for rib and metamaterial waveguides. The regions where the group velocity dispersion is positive correspond to the anomalous dispersion regime (see Figs. 2(a)-(c)). The crossings by zero of the integrated dispersion determine the position of the dispersive waves (see Figs. 2(d)-(f)). We considered a pump wavelength of 3.5 µm, in the anomalous dispersion regime, with a peak power of 1 kW. The metamaterial waveguide has a thickness of tSi = 700 nm, allowing fabrication with one lithography and one etching steps. The waveguide core has a width of Wc (see Fig. 1(a)). The metamaterial-grating cladding has a longitudinal period of Λ, with a gap length of lg and an anchoring arm length of la. The cladding width is set to Wclad = 5 µm to avoid leakage towards the lateral silicon slabs. For the rib waveguide, with a core width of WRib, we considered a core thickness of 700 nm and a slab thickness of 310 nm, similar to those used in ref [14]. Calculating the dispersion of the metamaterial waveguide requires finding the Bloch modes of a periodic waveguide. We do this using three-dimensional finite-difference time domain (FDTD) simulations. The dispersion of the rib waveguide can be calculated based on two-dimensional modal analysis, which we perform using finite-difference eigenmode (FDE) solver. In both cases we considered transverse-electric (TE) polarization.

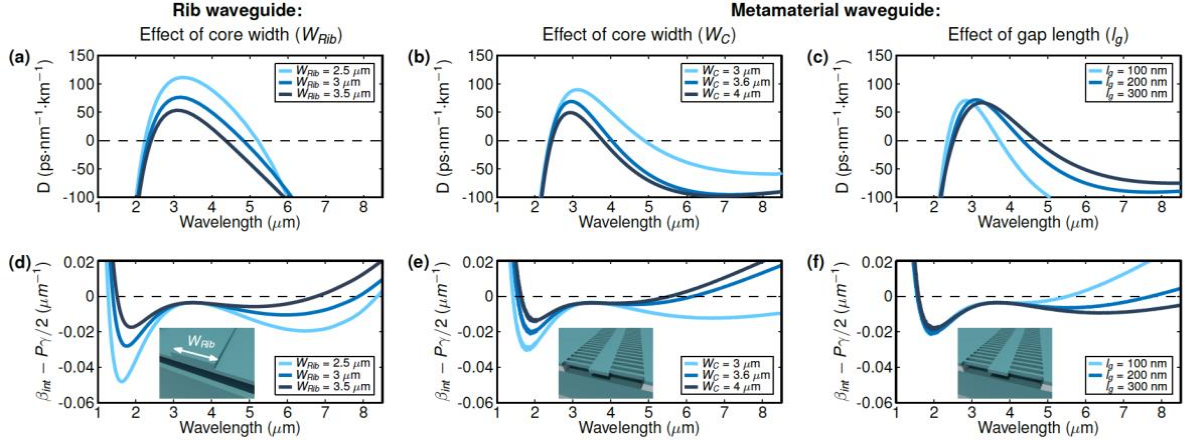

Fig. 2. Calculated group velocity dispersion (top panels) and integrated dispersion (bottom panels) for: (a),(d) Rib waveguide, considering different core widths $W_{Rib}$. Metamaterial waveguide, considering: (b),(e) variations in the core width ($W_C$) for $l_a$ = 150nm and $l_g$ = 150 nm, and (c),(f) variations in the gap length, $l_g$, for $W_C$ = 3.6µm and $l_a$ = 150 nm.

As shown in Fig. 2(d), variations of the rib core width ($W_{Rib}$) have a strong influence on the position of both dispersive waves, below and above the pump. Then, if we set a wavelength for one of the dispersive waves, the position of the other one is defined by the waveguide dispersion. This limited flexibility in controlling the position of the two dispersive waves hampers flexible shaping of the nonlinear generation spectrum. The width of the metamaterial waveguide core ($W_C$) also has a strong influence on the position of both dispersive waves (see Fig. 2(e)). Conversely, variations of the gap length ($l_g$) of the metamaterial waveguide have very little influence on the zero-crossing point at shorter wavelengths and strong effect on the zero-crossing point at longer wavelengths (see Fig. 2(f)). Thus, the metamaterial waveguide core width ($W_C$) and gap length ($l_g$) can be used to set the position of the dispersive waves at shorter and longer wavelengths, respectively. For example, choosing a waveguide width of $W_C$ = 3.6 µm, we set one dispersive wave at 1.55 µm wavelength. By changing $l_g$ we can tune the position of the other dispersive wave between 5 µm and 8 µm.

We study experimentally the effect of the metamaterial gap length ($l_g$) on the excitation of the dispersive waves that directly affects the bandwidth of the generated supercontinuum. For this, we fabricate three different waveguides with gap lengths of $l_g$ = 100 nm, $l_g$ = 200nm and $l_g$ = 300 nm, respectively. The three waveguides have a core width of $W_C$ = 3.6 µm and anchoring arm length of $l_a$ = 150 nm. The suspended waveguides were fabricated on a SOI wafer with 700-nm-thick silicon guiding layer and 3 µm-thick buried oxide (BOX) layer. The waveguide geometry was defined using one electron beam lithography and a single step of dry etching. The BOX layer was removed utilizing hydrofluoric (HF) acid vapor [30]. Figure 1(b) shows scanning microscope images of the fabricated waveguides.

The dispersive wave excitation and supercontinuum generation are experimentally characterized using the setup described in Fig. 3(a). The waveguides are pumped using a femtosecond laser coupled to an optical parametric amplifier (OPA). The resulting pulse has a repetition rate of 1 MHz and a pulse duration of 220 fs. Light is injected and extracted from the chip facet using 20 µm wide waveguides and aspherical ZnSe lenses (total insertion loss of ~24 dB, ~12 dB per facet). A polarization controller is used to inject TE polarized light into the chip. The chip output is analyzed using a monochromator and two photodetectors, an indium gallium arsenide photodetector for the near-IR wavelength range (1-2.4 µm wavelength) and a mercury cadmiun telluride photodetector for mid-IR wavelengths (2.4-13µm wavelength).

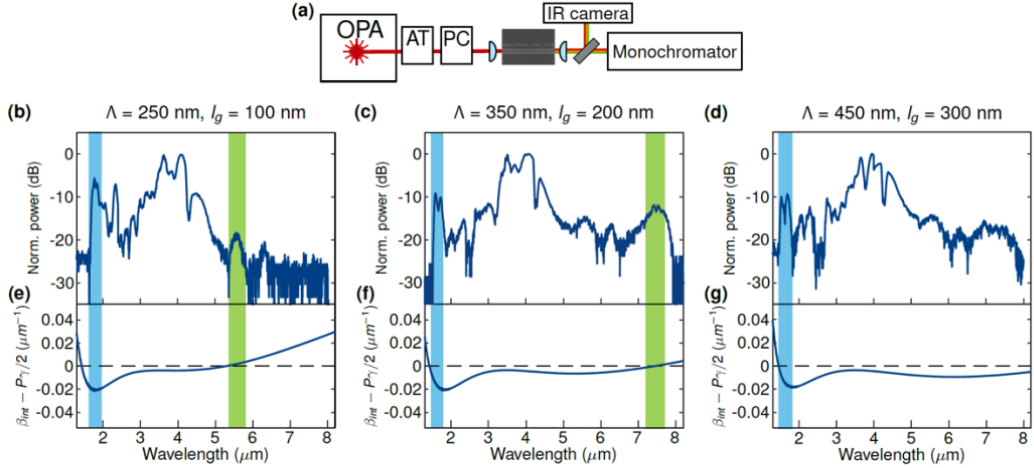

Fig. 3. (a) Experimental setup used to characterize supercontinuum generations. AT: attenuator. PC: polarization controller. Experimental supercontinuum characterization for pump wavelength of 3.5 μm wavelength and peak power coupled into the waveguide of 1 kW, for metamaterial waveguide with (b) $l_g$ = 100 nm, (c) $l_g$ = 200nm and (d) $l_g$ = 300 nm. Blue and green bars indicate the position of the measured dispersive wave below and above the pump wavelength, respectively. Calculated integrated dispersion for metamaterial waveguides with gap length of (e) $l_g$ = 100 nm, (f) $l_g$ = 200nm and (g) $l_g$ =300 nm. Crossings through zero indicate the predicted position of dispersive waves.

Figures 3(b)-(d) show the measured supercontinuum generation for the three 4-mm-long waveguides, with a pump wavelength of 3.5 μm and a measured average power near 3.5 mW. We estimate a peak power coupled into the waveguide of 1 kW, considering the measured average power before the chip, a pulse duration of 220 ps and coupling loss to the chip of 12 dB. The position of measured short-wavelength and long-wavelength dispersive waves are marked with blue and green bars, respectively. For comparison, in Figs. (e)-(f) we evaluate Eq. (1) for the three waveguides, considering a pump wavelength of 3.5 μm and a peak power of 1 kW. The crossings through zero indicate the calculated position of the dispersive waves. The dispersive wave positions obtained experimentally are in reasonably good agreement with the theoretical predictions via the waveguide integrated dispersion. The slight discrepancy in the short-wavelength dispersive wave position in the case of $l_g$ = 100nm could be attributed to fabrication imperfections that are accentuated for smaller gaps. We note that for $l_g$ = 300 nm, the estimated position of the long-wavelength dispersive wave falls outside the silicon transparency window (1.1-8 μm wavelength), precluding experimental observation. The gap length of the metamaterial geometry has a strong effect on the position of the dispersive wave at longer wavelengths, while exhibiting a minor impact on the position of the short-wavelength dispersive wave. For gap length of $l_g$ = 200 nm, the supercontinuum spectrum expands over 2.35 octaves across a wavelength range between 1.53 μm and 7.8 μm wavelength (-20 dB bandwidth). For this waveguide configuration ($W_C$ = 3.6 μm, Λ = 350 nm, and $l_g$ =200 nm) we measured propagation loss of 1-2 dB/cm in the 5-6 μm wavelength range, characterized using a quantum cascade laser with the cutback method with 6 different waveguides with lengths varying between 1.4 cm and 2.6 cm.

In summary, we propose and demonstrate a novel approach to control the excitation of dispersive waves on chip by exploiting silicon metamaterial waveguides. We show that judicious optimization of metamaterial waveguide geometry can yield an effective and independent control of the position of the short-wavelength and long-wavelength dispersive waves, independently. We experimentally demonstrate tuning of the long-wavelength dispersive wave between 5.5 μm and 7.5 μm wavelength, with negligible variation of the short-wavelength dispersive wave. We use this dispersion engineering flexibility to demonstrate a supercontinuum expanding over 2.35 octaves. This is, to the best of our knowledge, the widest supercontinuum generation in silicon and one of the widest demonstrations using integrated waveguides. These results open a promising route for the implementation of versatile nonlinear light broadening and frequency generation bridging telecom and mid-IR wavelengths using silicon photonics. We foresee that the metamaterial waveguide approach will foster the development of a new generation of high-performance nonlinear silicon photonic circuits exploiting flexible control of dispersive wave excitation for emerging applications in sensing metrology and communications.

**FUNDING**